\documentclass[aip,reprint]{revtex4-1}
\usepackage{graphicx}
\usepackage{dcolumn}
\usepackage{bm}
\usepackage[utf8]{inputenc}
\usepackage[T1]{fontenc}
\usepackage{xcolor}

\usepackage[upright]{fourier}
\usepackage{eso-pic}
\AddToShipoutPicture{
	\raisebox{2\baselineskip}{\makebox[\paperwidth]{\begin{minipage}[t][0.5cm][b]{21cm}\centering \scriptsize The following article has been accepted to the Journal of Vacuum Science \& Technology B. After it is published, it will be found at \texttt{https://avs.scitation.org/journal/jvb}.\end{minipage}
}}}

\begin{document}
\renewcommand{\figurename}{{\color{black}Fig.}}
\preprint{}

\title[Direct Laser Writing of Birefringent Photonic Crystals]{Direct Laser Writing of Birefringent Photonic Crystals for the Infrared Spectral Range}

\author{M. Lata}
 \email{mlata@uncc.edu}
 \affiliation{Department of Physics and Optical Science, University of North Carolina at Charlotte, 9201 University City Blvd, Charlotte, NC, 28223}
\author{Y. Li}
 \affiliation{Department of Physics and Optical Science, University of North Carolina at Charlotte, 9201 University City Blvd, Charlotte, NC, 28223}
\author{S. Park}
 \affiliation{Department of Physics and Optical Science, University of North Carolina at Charlotte, 9201 University City Blvd, Charlotte, NC, 28223}
\author{M. J. McLamb}
 \affiliation{Department of Physics and Optical Science, University of North Carolina at Charlotte, 9201 University City Blvd, Charlotte, NC, 28223}
\author{T. Hofmann}
 \affiliation{Department of Physics and Optical Science, University of North Carolina at Charlotte, 9201 University City Blvd, Charlotte, NC, 28223}

\begin{abstract}
Infrared optical {\color{black}photonic crystals} fabricated using direct laser writing, which is based on the two-photon polymerization of suitable monomers, have received substantial interest since the emergence of this process. Two-photon polymerization could be a disruptive technology for the fabrication of all-dielectric {\color{black}photonic crystals} in the infrared spectral range, as it allows the synthesis of large scale arrays of uniform structures with arbitrary geometries and arrangements. However, all-dielectric photonic crystals that provide birefringent optical responses in the infrared spectral range have not yet been demonstrated using direct laser writing techniques. Here we explore the form birefringence observed in photonic crystals composed of arrays of subwavelength-sized slanted polymer microwires. The photonic crystals investigated here were fabricated in a single fabrication step using direct laser writing of an infrared transparent photoresist (IP-Dip). A strong contrast of the cross-polarized reflectance of photonic crystals as a function of the in-plane orientation is observed in the mid-infrared spectral range at {\color{black}$\lambda$ $\approx$ 6.5~$\mu$m}. This observation is indicative of an anisotropic optical behavior. Finite element based techniques corroborate the experimentally observed responses {\color{black} qualitatively}. 
\end{abstract}
\maketitle

\section{Introduction}
 {\color{black}Photonic crystals} have received increased attention in recent years for optical applications ranging from chiral optical filters to super resolution.\cite{RN73} Many {\color{black}photonic crystals} are fabricated from metals to take advantage of plasmonic resonances.\cite{RN73} However, these {\color{black}photonic crystals} tend to suffer from Ohmic losses.\cite{RN65} All-dielectric {\color{black}photonic crystals} have therefore been considered in order to eliminate the inherent Ohmic losses observed in their metal-based counterparts.\cite{RN70} Recently, all-dielectric {\color{black}photonic crystals} have been demonstrated to display a range of interesting optical properties such as perfect reflectance and strong, lossless resonant responses through trapped magnetic modes, for instance.\cite{RN71,RN64} 
 
 {\color{black}Photonic crystals} derive their unique optical properties largely from the geometry of their subwavelength-sized building blocks. Therefore, sophisticated fabrication techniques are required to synthesize the subwavelength-sized elements with desired geometries and arrangements.\cite{RN73} Among applicable fabrication techniques, direct laser writing by two-photon polymerization of suitable monomers enables the fabrication of arbitrary geometries with critical features on the micro- and even nanometer-scale.\cite{RN60,RN61} Thus, direct laser writing by two-photon polymerization has received substantial interest as an efficient avenue to fabricate all-dielectric {\color{black}photonic crystals} for the infrared spectral range. 
 
 We have demonstrated that direct laser writing allows the fabrication of structured surfaces to reduce Fresnel reflection loss at infrared wavelengths.\cite{RN75} Furthermore, all-dielectric, two-dimensional photonic crystals for optical filters with high spectral contrast in the infrared spectral range have been successfully fabricated using this process.\cite{RN74} 
 
 Despite these advances in the fabrication and characterization of all-dielectric {\color{black}photonic crystals}, information on the anisotropic optical responses of this material class is still lacking. In particular, geometries of all-dielectric photonic crystals composed of regular arrangements of  subwavelength-sized slanted wires have not been explored in the infrared spectral range yet. This is in contrast to the visible and near-infrared spectral range, where such geometries, fabricated using UHV deposition techniques, have been demonstrated to exhibit strong form-induced birefringence and dichroism.\cite{RN67,RN68,RN69}      
 
 Here, we report on the fabrication of all-dielectric birefringent photonic crystals composed of slanted microwire arrays using direct laser writing of a photoresist. The fabricated photonic crystals were investigated using  polarization-sensitive infrared microscopy. A distinct contrast was observed between photonic crystals with different azimuthal orientations. In addition, complementary finite element based calculations were performed that corroborate these experimental findings {\color{black} qualitatively}.
 
 \begin{figure}[b!]
	\centering			
	\includegraphics[width=0.7\linewidth,keepaspectratio]{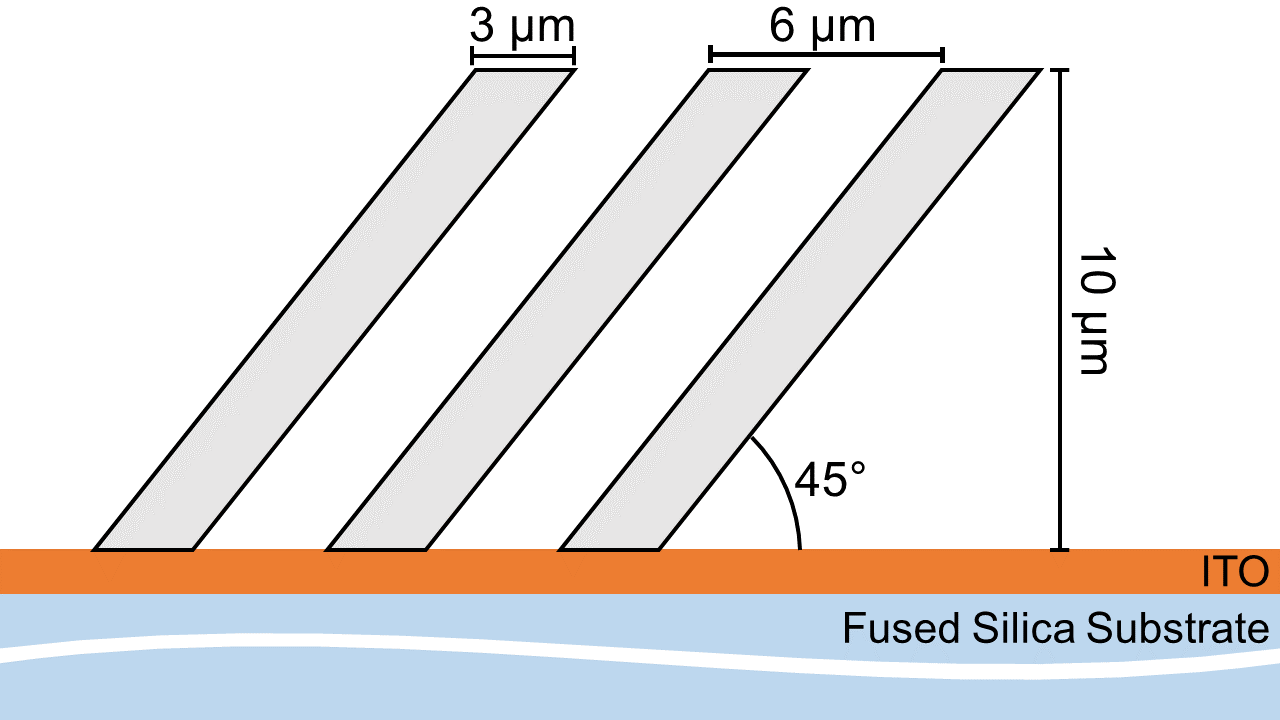}
	\caption{Side view of the birefringent photonic crystal design. The slanted microwires have a square cross-section and are arranged in a square array of approximately 50 $\mu$m $\times$ 50 $\mu$m.}
	\centering		
	\label{fig:design}
\end{figure}

\section{Design \& Fabrication}
The all-dielectric birefringent photonic crystals investigated here are composed of slanted microwires as shown in Fig.~\ref{fig:design}. The slanted microwires have a square base with nominal dimensions of 3$\times$3 $\mu$m. The slanting angle of the microwires is 45$^{\circ}$ with respect to the plane of the substrate and the microwires extend 10~$\mu$m vertically from the substrate. The microwires are arranged periodically over a square of approximately 50 $\mu$m $\times$ 50 $\mu$m with a center distance of 6 $\mu$m. 

The photonic crystals are designed to operate in the infrared spectral range. Therefore, the dimensions of the slanted wires were chosen to maintain a subwavelength regime for infrared wavelengths, while being sufficiently large to allow accurate fabrication using the direct laser writing process. 

In order to conveniently test the polarization response of the photonic crystals in the infrared spectral range, four photonic crystals with different azimuthal orientations of the microwire slanting plane ranging from $\varphi=0^{\circ}$ to 135$^{\circ}$ in steps of 45$^{\circ}$ were arranged in a pixel-like pattern. 

The photonic crystals shown in Fig.~\ref{fig:design} were  fabricated using a commercially available 3D direct laser writing system (Photonic Professional GT, Nanoscribe, GmbH).
IP-Dip was selected as a suitable photoresist, as it allows the fabrication of photonic crystals with $\mu$m- to nm-scale critical dimensions.\cite{RN75} ITO-coated fused silica was selected as a substrate material. The all-dielectric photonic crystals were fabricated in a single 3D direct laser writing step using a 25$\times$ objective. This approach allows the fabrication of the photonic crystals by steering the laser beam with the galvanometer scanner of the 3D direct laser writing system. By using only the galvanometer scanner, the need for any mechanical translation of the sample stage is omitted. This eliminates possible stitching and over-exposure errors that may be induced by the mechanical translation of the sample.\cite{Oakdale2017direct}

\begin{figure}[t]
	\centering			
	\includegraphics[width=0.65\linewidth,keepaspectratio]{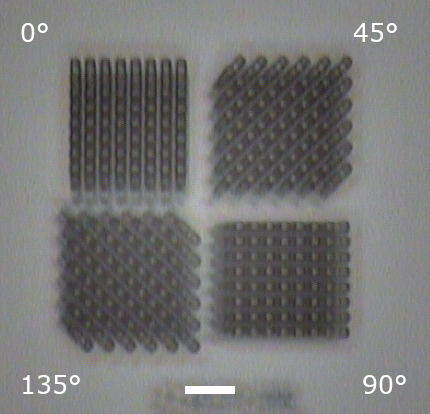}
	\caption{Top-view optical micrograph of the fabricated birefringent photonic crystals. The individual photonic crystals with azimuthal orientations ranging from 0$^{\circ}$ to 135$^{\circ}$ are clearly distinguishable. The scale bar indicates 20~$\mu$m.}
	\label{fig:micrograph}
	\centering			
\end{figure}

\begin{figure}[t]
	\centering			
	\includegraphics[width=0.65\linewidth,keepaspectratio, trim=0 0 0 0,clip]{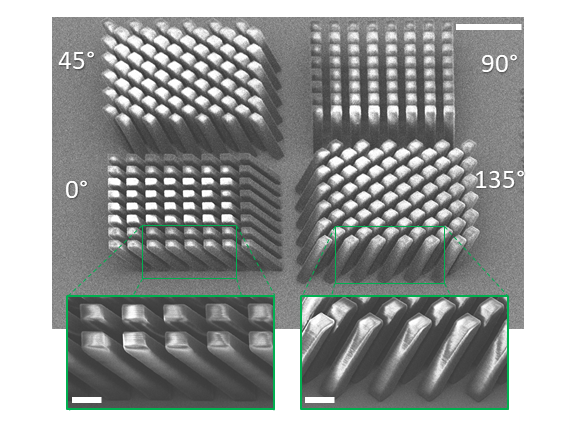}
	\caption{\color{black}Tilted (approx.~50$^{\circ}$) top-view SEM micrograph of fabricated birefringent photonic crystals. As in Fig.~2, the individual photonic crystals with azimuthal orientations ranging from 0$^{\circ}$ to 135$^{\circ}$ are clearly distinguishable. The geometry of the fabricated wires are in very good agreement with the nominal design. The scale bars indicate 20~$\mu$m and 5~$\mu$m, for the top panel and the insets, respectively.}
	\label{fig:SEM}
	\centering			
\end{figure}

After the fabrication, any unpolymerized monomer was removed by immersing the samples in propylene glycol monomethyl ether acetate (PGMEA, Baker 220) for 20~min. Subsequently, the samples were immersed in 99.99\% isopropyl alcohol for 2~min. Finally, the remaining isopropyl alcohol was evaporated at room temperature. 

An optical micrograph of the fabricated photonic crystals is shown in Fig.~\ref{fig:micrograph}. The four photonic crystals with azimuthal orientations ranging from $\varphi=0^{\circ}$ to 135$^{\circ}$ are clearly distinguishable and the fabricated wires appear to be true to the designed form. {\color{black} In addition, a SEM micrograph is depicted in Fig.~\ref{fig:SEM}, clearly showing an excellent agreement between the fabricated structure geometries and the nominal design shown in Fig.~\ref{fig:design}.}

\section{Experimental Results \& Finite Element Modeling}
\begin{figure}[t]
	\centering			
	\includegraphics[width=0.6\linewidth,keepaspectratio]{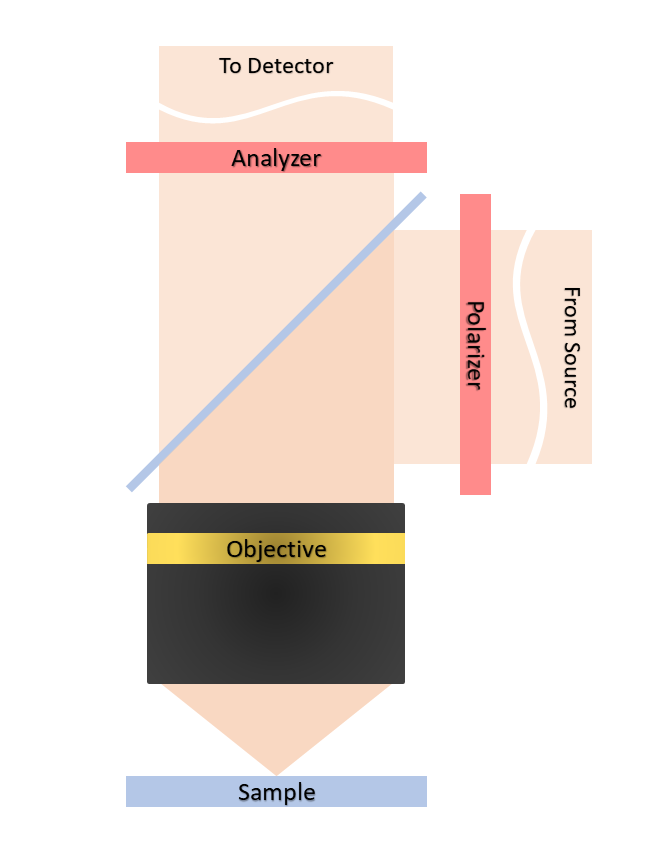}
	\caption{ \color{black} Simplified optical beam path for the Bruker Hyperion 3000 microscope utilized for the reflectance measurements shown in Fig.~6. The collimated illumination from the globar source is polarized before being focused by a Cassegrain objective onto the sample. The reflected light is then collected and re-collimated by the same objective and sent through an analyzer oriented orthogonally to the input polarizer before being sent to the detector.}
	\label{fig:beampath}
\end{figure} 
The cross-polarized reflectance of these photonic crystals was simulated using a commercial finite element modeling software package (COMSOL Multiphysics) in order to determine their expected infrared optical responses. The dielectric function of IP-Dip, which was previously determined in the infrared spectral range by spectroscopic ellipsometry, was used for these model calculations.\cite{RN66} {\color{black}The structures were imported into the finite element software and arranged onto a perfectly reflecting substrate, and a frequency domain plane-wave reflection simulation was performed.} 

\begin{figure}[b]
 	\centering			
 	\includegraphics[width=0.7\linewidth,keepaspectratio]{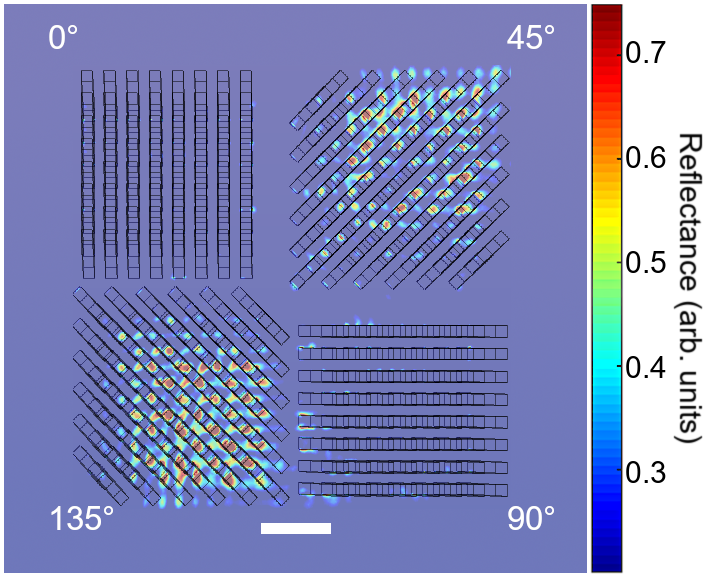}
 	\caption{Finite element model-calculated reflectance image under cross-polarized illumination obtained for an array of all-dielectric photonic crystals at normal incidence and {\color{black}$\lambda$ $\approx$ 6.5 $\mu$m}. Four photonic crystals with azimuthal orientations ranging from $\varphi=0^{\circ}$ to 135$^{\circ}$ are shown here. An infinite plane wave was assumed for the model calculation. The scale bar indicates 20~$\mu$m.}
 	\label{fig:comsol}
 \end{figure}

Fig.~\ref{fig:comsol} shows the model calculated relative intensity image under cross-polarized illumination obtained for an array of four photonic crystals at normal incidence and {\color{black}$\lambda$ $\approx$ 6.5 $\mu$m}. {\color{black}The cross polarized response was found by using a linearly polarized incident plane wave and then measuring the reflected intensity of the light polarized orthogonally to the input wave.}  A strong contrast between the crystals with azimuthal orientations of $\varphi=0^{\circ}$ and $90^{\circ}$, and $\varphi=45^{\circ}$ and $135^{\circ}$, respectively, can be observed. For the photonic crystals for which the optical axis, which is parallel to the slanted wire elements, is oriented obliquely with respect to the input polarization ($\varphi=45^{\circ}$ and $135^{\circ}$), the cross-polarized reflectance is the largest. For the photonic crystals for which the optical axis is oriented parallel or perpendicular to the input polarization ($\varphi=0^{\circ}$ and $90^{\circ}$), the cross-polarized reflectance is vanishing. This optical response is indicative of a significant polarization mode conversion due to the form-induced birefringence of the investigated photonic crystals in the infrared spectral range.
The cross-polarized reflectance of the photonic crystals was measured at {\color{black}$\lambda$ $\approx$ 6.5 $\mu$m} using a Bruker Hyperion 3000 infrared microscope in combination with a Bruker Vertex 70 FTIR spectrometer. {\color{black}The beam path for this microscope when measuring in reflectance is outlined in Fig.~\ref{fig:beampath}.} The microscope is equipped with KRS-5 wire grid polarizers in the input and output beam path in order to allow polarization sensitive measurements. A liquid nitrogen cooled Mercury-Cadmium-Telluride focal plane array with 64~$\times$~64 pixels was used as a detector {\color{black}and a SiC globar was used as a source}. In combination with a 15$\times$ Cassegrain objective, the instrument allows the acquisition of hyperspectral infrared images with a lateral resolution of 2.7~$\mu$m.

\begin{figure}[b]
	\centering			
	\includegraphics[width=0.7\linewidth,keepaspectratio]{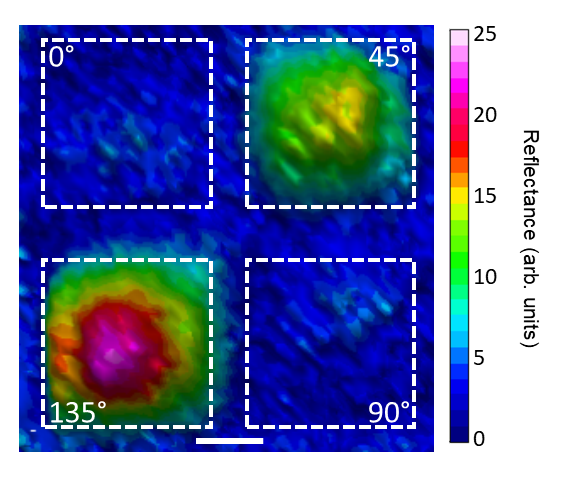}
	\caption{Cross-polarized reflectance image of the fabricated birefringent photonic crystals obtained at normal incidence and {\color{black}$\lambda$ $\approx$ 6.5 $\mu$m}. The dotted boxes illustrate the surface area of the individual photonic crystals with an azimuthal orientation from $\varphi=0^{\circ}$ to $\varphi=135^{\circ}$ with respect to the incident polarization direction. The scale bar indicates 20~$\mu$m.}
	\label{fig:reflectance}
\end{figure}

Fig.~\ref{fig:reflectance} shows a cross-polarized reflectance image of the photonic crystal array obtained at {\color{black}$\lambda$ $\approx$ 6.5 $\mu$m} and normal incidence. The measured cross-polarized reflectance shows a similar contrast as observed in the finite element model-calculated reflectance shown in Fig.~\ref{fig:comsol}. The cross-polarized reflectance is maximal for the photonic crystals with azimuthal orientations of $\varphi=45^{\circ}$ and $\varphi=135^{\circ}$ for which the optical axis of the photonic crystals is obliquely oriented with respect to the input polarization. For the photonic crystals with azimuthal orientations of $\varphi=0^{\circ}$ and $\varphi=90^{\circ}$, the reflectance is vanishing. This behavior is in good qualitative agreement with the observations from the finite element based model calculations and confirms experimentally the birefringence of the investigated all-dielectric photonic crystals. {\color{black} The calculated near field response depicted in Fig.~\ref{fig:comsol} shows distinct intensity fluctuations under cross-polarized illumination. The experimentally obtained response (Fig.~\ref{fig:reflectance}) does not exhibit these fluctuations across the surface of the individual crystals, as the far field is being recorded.}
Careful inspection of Fig.~\ref{fig:reflectance} shows that there is a considerable difference between the reflectance obtained for the photonic crystals with $\varphi=45^{\circ}$ and $\varphi=135^{\circ}$. We tentatively attribute this intensity difference to a biaxial optical response which would be consistent with observations for slanted columnar thin films reported in the visible spectral range.\cite{RN69, Hodgkinson1997} In addition, it can be noticed that the intensity distribution across the surface of the photonic crystals is nonuniform. These apparent boundary effects are currently under investigation and isotropic support structures designed to mitigate their impact are being explored.


\section{Conclusions}
All-dielectric photonic crystals comprised of slanted microwires were successfully fabricated using direct laser writing from an infrared transparent photoresist. Experimentally obtained cross-polarized reflectance images exhibit a strong contrast between photonic crystals as a function of the azimuthal orientation. Photonic crystals obliquely oriented with respect to the direction of the input polarization show a strong reflectance. The cross-polarized reflectance of the crystals oriented perpendicular or parallel to the input polarization is vanishing. The experimental observations are found to be in qualitative agreement with finite element based model calculations of the cross-polarized normal incidence reflectance. The optical response is indicative of a birefringent behavior which is expected for the photonic crystals composed of slanted microwires. The difference between the cross-polarized reflectance of photonic crystals with oblique azimuthal orientation is tentatively interpreted as an indication of a biaxial response. This would be consistent with experimental findings reported in the literature for slanted columnar thin films investigated in the visible spectral range. We have also noticed boundary effects, which could hinder the application of these materials for pixel-based polarization filters for wavefront-splitting infrared polarimeters, for instance.
 
\section*{\label{acknowledgement}Acknowledgement}
The authors are grateful for support from the National Science Foundation (1624572) within the I/UCRC Center for Metamaterials, the Swedish Agency for Innovation Systems (2014-04712), and Department of Physics and Optical Science of the University of North Carolina at Charlotte.


\end{document}